\documentclass{webofc}
\usepackage[varg]{txfonts}   
\usepackage{wrapfig}              
\usepackage{adjustbox}          

\begin{document}
\title{Nuclear Fission:  from more phenomenology and adjusted parameters to more fundamental theory and increased predictive power  }

\author{ \firstname{Aurel}       \lastname{Bulgac}         \inst{1}     \fnsep\thanks{\email{bulgac@uw.edu}} \and
              \firstname{Shi}          \lastname{Jin}               \inst{1}     \fnsep \and
              \firstname{Piotr}        \lastname{Magierski}    \inst{2,1}      \fnsep \and
              \firstname{Kenneth}  \lastname{Roche}          \inst{3,1}     \fnsep  \and
              \firstname{Nicolas}    \lastname{Schunck}      \inst{4}     \fnsep \and
              \firstname{Ionel}        \lastname{Stetcu}          \inst{5}     \fnsep
}

\institute{  Department of Physics, University of Washington., Seattle, WA 98195-1560, US
\and
                 Faculty of Physics, Warsaw University of Technology, ulica Koszykowa 75, 00-662 Warsaw, POLAND
\and
                 Pacific Northwest National Laboratory, Richland, WA 99352, USA
\and
                 Nuclear and Chemical Science Division, Lawrence Livermore National Laboratory, Livermore, CA 94551, USA
\and
                 Theoretical Division, Los Alamos National Laboratory, Los Alamos, NM 87545, USA          }

\abstract{ Two major recent developments in theory and computational resources created
the favorable conditions for achieving a microscopic description of nuclear fission
almost eighty years after its discovery in 1939 by Hahn and Strassmann~\cite{Natur_1939r}.
The first major development was in theory, the extension of the Time-Dependent Density Functional
Theory (TDDFT)~\cite{hk,ks,monograph1,monograph2} to superfluid
fermion systems~\cite{ARNPS__2013}. The second development was in computing, the emergence of
powerful enough supercomputers capable of solving the complex systems of equations
describing the time evolution in three dimensions without any restrictions
of hundreds of strongly interacting nucleons.
Even though the available nuclear energy density functionals (NEDFs) are phenomenological still, their
accuracy is improving steadily and the prospects of being able to perform calculations of the
nuclear fission dynamics and to predict many properties of the fission fragments, otherwise not
possible to extract from experiments,  are within reach, all without making recourse anymore to
uncontrollable assumptions and simplified phenomenological models.
}
\maketitle

Meitner and Frisch~\cite{Nature_1939r} and Bohr and Wheeler~\cite{Nature_1939r1,PhysRev_1939r}
interpreted the neutron induced fission of uranium observed
by Hahn and Strassmann~\cite{Natur_1939r} as the Coulomb-driven division of a classically charged liquid drop
in competition with the surface tension of the liquid drop and they
obtained a good back-of-the-envelope estimated of the total kinetic energy (TKE)
of the fission fragments (FFs). Unlike the $\alpha$-decay and spontaneous fission
observed later~\cite{Flerov},  induced fission occurs above the barrier, therefore is
to a large extent a classical process. The impinging neutron on uranium leads to
the formation of an excited state, the compound nucleus,
in a region of the nuclear spectrum where the level density is extremely high
and the nucleus is thus relatively hot.
The nucleus evolves for a very long time from its ground state shape towards the top of the fission,
where the Coulomb repulsion overcomes surface
tension and the nucleus starts its descent to scission, where the two FFs are formed.
The top of the fission barrier is energetically barely below the neutron threshold. Although
the nucleus is already quite elongated at this point, it has relatively little
intrinsic excitation energy and is relatively cold.
\begin{wrapfigure}{r}{0.35\textwidth}
\includegraphics[clip,width=0.35\textwidth]{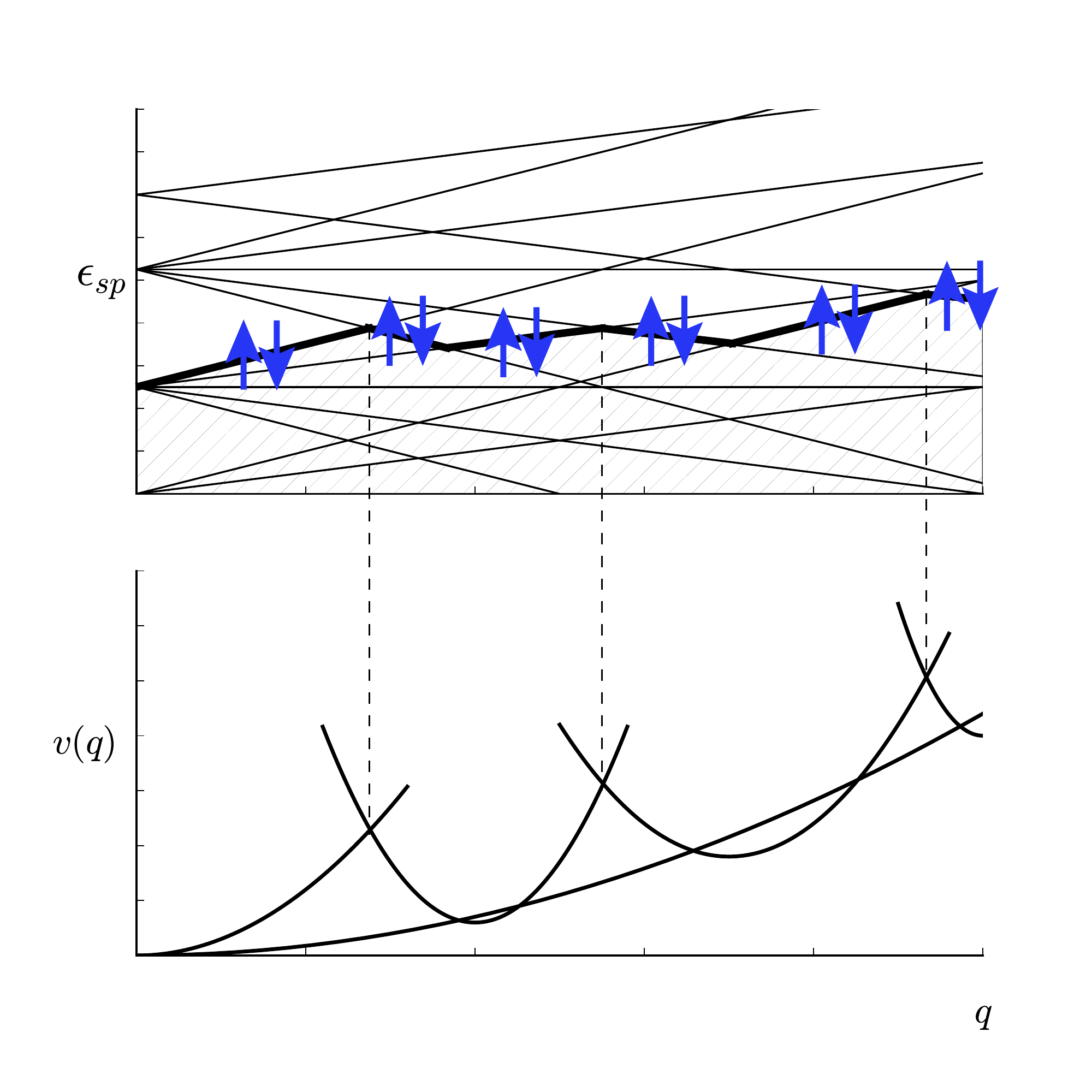}
\caption{ \label{fig:ab2}    The qualitative evolution of the single-particle levels (upper panel) and
of the total nuclear energy (lower panel) as a function of
nuclear deformation~\cite{PhysRev_1953r, Bertsch:1980}. The Fermi level is shown with a thick line.}
\end{wrapfigure}
It was soon realized that in nuclei nucleons form shells and behave
in many instances as independent particles, like electrons in atoms.
In other words nucleons live on quantized
orbits~\cite{Mayer,Jensen}, and the spin-orbit interaction plays
a critical role in the formation of the nuclear shells and in the formation of the FFs~\cite{Meitner}.  Hill and
Wheeler~\cite{PhysRev_1953r} discussed how the liquid drop deformation energy emerges from a quantum
mechanical approach based on considering the quantized single-particle
motion of nucleons in a slowly deforming potential well. In their approach, the liquid
drop potential deformation energy was, in the first
approximation, an envelope of many intersecting parabolas caused by
single-particle level crossings, see Figure \ref{fig:ab2}. At
single-particle level crossings of the last occupied level, nucleons jump from one level to another, in order to maintain the
sphericity of the Fermi sphere. Without such a redistribution of nucleons
at the Fermi level, the Fermi sphere would become oblate when a nucleus deforms on the way to
scission into two fragments, while the
spatial shape of the nucleus would become ever more prolate. This would lead to a
volume excitation energy of the nucleus. This is inconsistent with the fact that nuclei
are saturating systems with a surface tension that can only
deform by changing the shape of their surface while maintaining
their volume constant. Hence, only the Coulomb and the surface
contributions to the total energy changes. Each single-particle level is typically double degenerate, due to
Kramers degeneracy, and nucleons have to jump in pairs,
otherwise the nuclear shape evolution towards scission would be
hindered~\cite{Bertsch:1980,PhysRevLett_1997r}. In this case, nucleons would
follow diabatic levels instead of adiabatic ones, as was indeed established
in recent TDHF simulations of fission dynamics~\cite{arxiv_2015r,arxiv_2015r1}. Pairing correlations,
while relatively weak in nuclei, is very effective
at promoting simultaneously two nucleons from time-reverse orbits into
other time-reverse orbits and thus it greatly facilitates the
evolution of the nuclear shape towards scission.

It was established later that single-particle level bunching exists in
nuclear systems not only in the case of spherical nuclei (as in the
case of atoms), but also in deformed and highly deformed nuclei. At
first this phenomenon was experimentally observed in the case of
fission isomers at very large
elongations~\cite{Polikanov,RMP_1972r,RMP_1980r} and subsequently in
the case of superdeformed nuclei~\cite{Twin}.  The existence of
nucleonic shells at large deformations results in a potential energy
deformation surface with significant maxima and minima, which are
otherwise absent in the case of a classical charged liquid drop. The
aforementioned level crossings cause discontinuities in the potential energy surface.
Some of these discontinuities can be removed when pairing
correlations are present, since their effect is to avoid level crossings.

Overall fission dynamics is a very complex process, which still
did not reach a fully microscopic description~\cite{arxiv_1511r} in
spite of almost eight decades of effort.
The formation of FFs was and still is one of the most debated topics in
phenomenology of nuclear fission. In the absence of a  microscopic description and
because of the enormous challenges arising from solving the necessarily time-dependent
Schr\"{o}dinger equation for hundreds of strongly interacting nucleons, many
phenomenological models, both quantum and classical in nature,
have been suggested over the years. Unfortunately most of them
lack a solid microscopic underpinning and for that reason their predictive power is
typically limited to the processes used to construct these models.

The bulk behavior of the deformation potential energy surface is determined
by the surface tension
and Coulomb energy, which are rather well-known and understood,
and they are the main driving force leading to
fission as in liquid drop models.  The underlying shell structure imprints hills and
valleys on the
overall charged liquid drop energy landscape~\cite{RMP_1972r,RMP_1980r}. On the way to scission
configuration, nucleons have to go through a large number of
redistributions at the single-particle levels crossing around the
Fermi level, in order to maintain the spherical symmetry of their
local momentum distribution.
\begin{figure}[!ht]
\includegraphics[clip,width=0.45\textwidth]{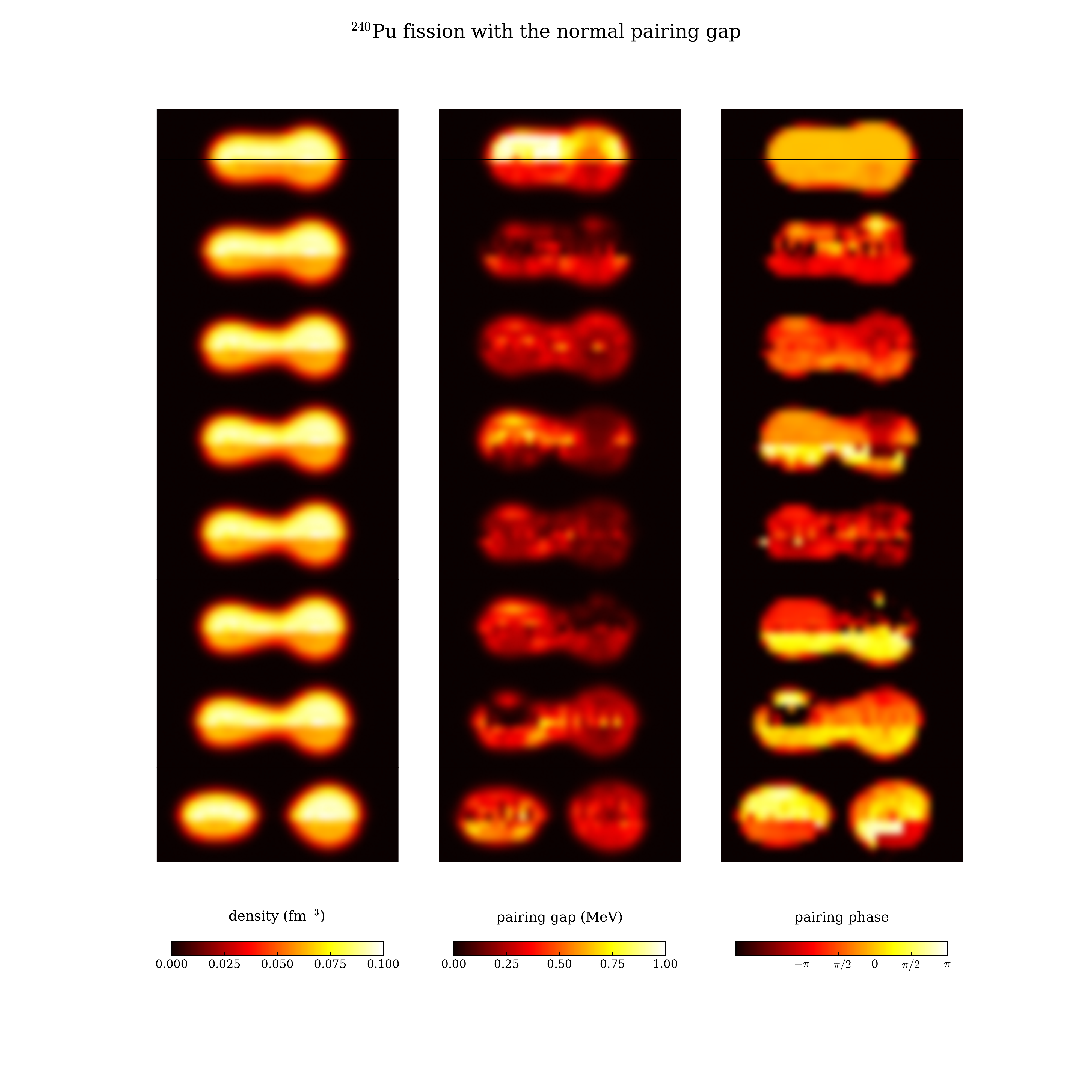}
\includegraphics[clip,width=0.45\textwidth]{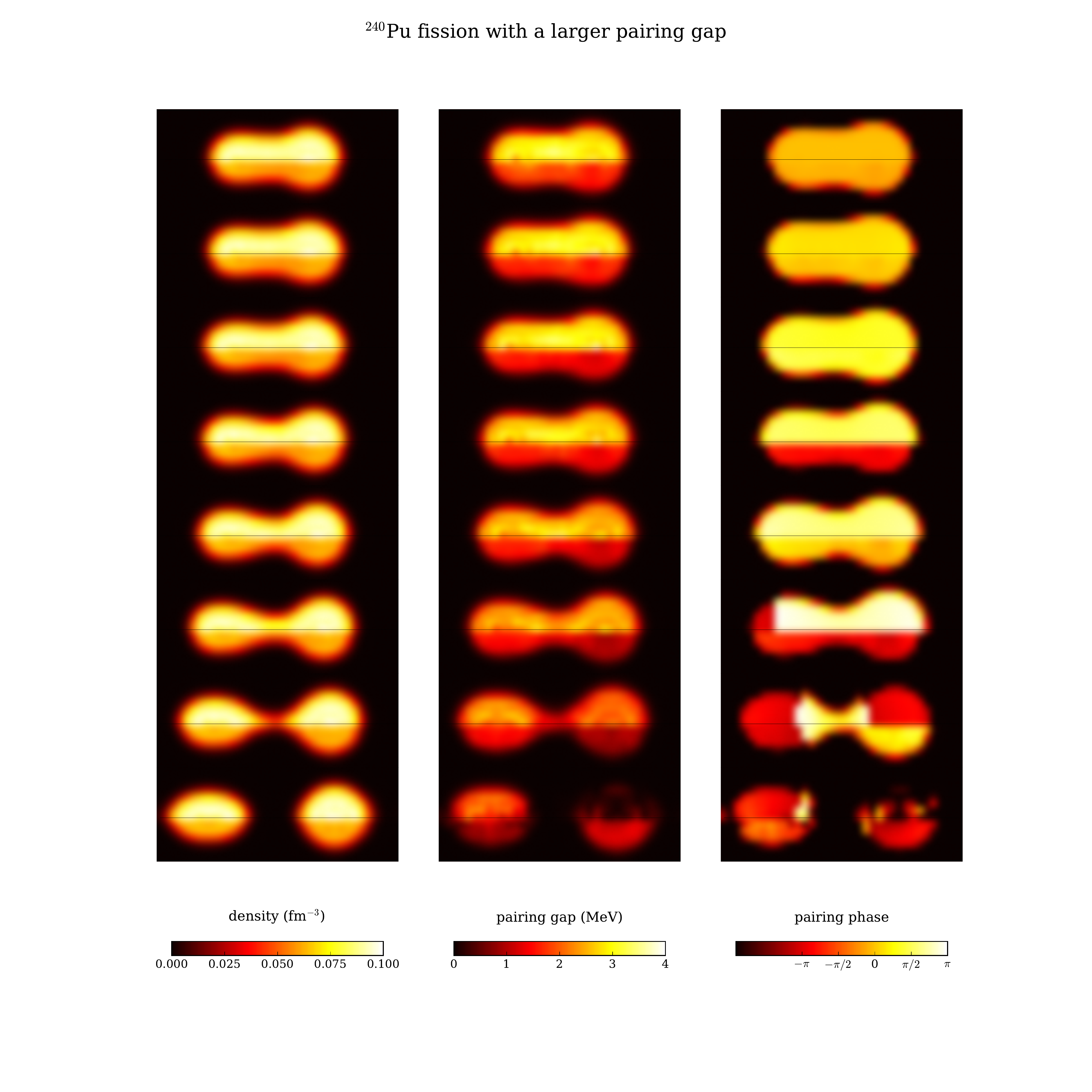}
\caption{ \label{fig:ab34}   In the left panel the induced fission of $^{240}$Pu with normal
pairing strength last up to 14,000 fm/c ($\approx 47\times 10^{-21}$ s) from saddle-to-scission. The
columns show sequential frames of the density (first column), the
magnitude of the pairing field (second column), and the phase of the
pairing field (third column).  In each frame the upper/lower part of
each frame shows the neutron/proton density, the magnitude of
neutron/proton pairing fields, and of the phase of the pairing field
respectively~\cite{Bulgac:2016}. At scission the heavy fragment is on
the right and emerges almost spherical, while the light fragment is
highly deformed with the ratio of the axes $\approx 3/2$.
In the right panel the induced fission of $^{240}$Pu with
enhanced pairing strength last about 1,400 fm/c from
saddle-to-scission,  in a time less than the time
between the consecutive frames in the left panels.}
\end{figure}
Overall, the
deformation potential energy surface acquires a profile somewhat
similar to that of an uneven mountain, with little hills and valleys
and covered by trees, and the evolution of the nuclear shape is in the
end similar to the erratic motion of a pinball, not straight down the
hill, but rather left and right, bouncing (mostly elastically) from
the many obstacles on the way to the bottom of the valley, where the
pinball breaks up. In the last stages of this complex
evolution, the independent character of the nucleons inside the nucleus
plays again a critical role, since magic closed shells control the nuclear
shape evolution and the splitting leading to the formation of FFs.

Since fission dynamics is a truly non-stationary phenomenon an
extension of the TDDFT~\cite{hk,ks,monograph1,monograph2} was needed~\cite{ARNPS__2013}
and this approach was dubbed the Time-Dependent Superfluid Local
Density Approximation (TDSLDA).
In the case of many strongly
interacting particles the Schr\"{o}dinger equation has to be replaced with the
TDDFT, which is mathematically equivalent to the Schr\"{o}dinger equation for one body observables and the only theoretical framework suitable to describe the structure and dynamics of
quantum many-body systems, if the corresponding energy density functional (the equivalent of the potential in the Schr\"{o}dinger equation) is known in sufficient detail.
TDSLDA allows the evaluation of  FFs properties, which are otherwise impossible to either measure or
compute, such as FFs excitation energies.
As our simulations within TDSLDA  demonstrate~\cite{Bulgac:2016}, see Figure~\ref{fig:ab34},
the nucleus separates typically into two FFs, one bigger
and the other somewhat smaller. In our calculations, we used in the particle-hole channel a
phenomenological NEDF~\cite{NuclPhys_1998},  complemented
with pairing correlations described as in Refs.~\cite{PRL__2002,PRL__2003a}.
The larger fragment fragment, which has
properties very similar to the energetically very stable double-magic
$^{132}$Sn, emerges with an almost spherical shape, while the lighter
fragment emerges in an elongated configuration, with a ratio
of the major to minor axes close to 3/2.
To much of our surprise, the properties of the FFs in these simulations~\cite{Bulgac:2016} are
very close experimental measurements, even though no effort was put into reproducing observations.
We attribute this to the fact that the NEDF reproduces bulk nuclear properties
relatively well. The biggest surprise, however, was the observation that the time from
saddle-to-scission was about an order of magnitude larger than ever estimated previously in the literature.
This was ulteriorly confirmed by independent simulations~\cite{Tanimura}.

The right panel of Figure~\ref{fig:ab34} serves as an illustration of the crucial role
played by pairing correlations in fission dynamics. In the left panel one can see that the
pairing field on the way from saddle-to-scission does fluctuate noticeably in magnitude and
phase. Strictly speaking, therefore, the pairing field during its time evolution stops being a
superfluid condensate of Cooper pairs, which otherwise would show a uniform phase. By
increasing artificially the strength of the pairing field, we observe that fission dynamics
is accelerated significantly, see right panel in Figure~\ref{fig:ab34}. In that case, the pairing field shows
the expected characteristics of a slowly-evolving superfluid condensate,
the nuclear fluid behaving almost like a perfect or ideal fluid, where dissipation is
practically absent, and the two FFs emerge macroscopically entangled now~\cite{Bulgac:2017}.

At FUSION17 workshop W. Mittig~\cite{mittig}
discussed the very intriguing possibility of the existence of long-lived nuclear systems
with very large charges, $Z\approx 200$  or even higher. One can contemplate the collision of two
heavy nuclei with an energy just at the Coulomb barrier, when fusion is very likely to occur. As in
the case of fission dynamics, immediately after fusion many collective degrees of freedom of the
combined system are likely to get excited, leading to a sharing of the energy from the incoming
collision channel. This combined nuclear system might thus survive
for a relatively long time, but eventually the system
will decay into two almost equal fragments, as the combined system would likely equilibrate,
even in the case of asymmetric collisions.

\begin{wrapfigure}{r}{0.36\textwidth}
\includegraphics[clip,width=0.36\textwidth]{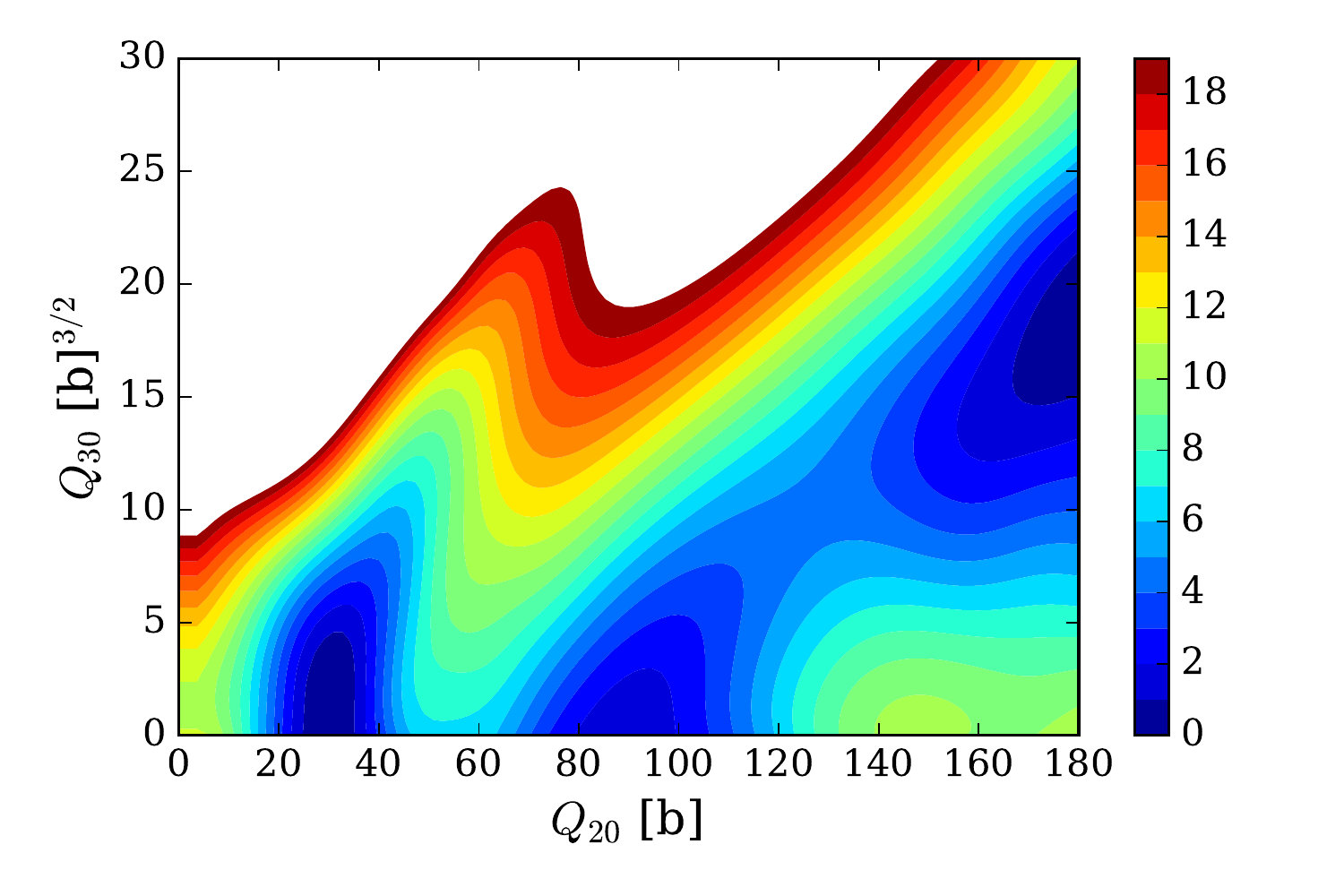}
\caption{ \label{fig:ab5}   The potential energy surface of $^{240}$Pu evaluated with
the new energy density functional SeaLL1~\cite{NEDF}.}
\end{wrapfigure}
A major limitation of the DFT is a lack of a method to evaluate two-body observables, such
as the width of the FFs mass and charge yields and of the total TKE
of the FFs. Within the TDHF approach one can use the Balian and
V\'{e}n\'{e}roni~\cite{Balian} prescription or the more involved stochastic
mean-field model introduced by Ayik~\cite{Ayik}, which reproduces the widths
evaluated within the Balian and V\'{e}n\'{e}roni procedure. Ayik's model is
computationally more costly, but is the first attempt to describe
quantum mechanically within TDDFT the full distribution
of various observables, which in case of fission would amount to the FFs mass,
charge yields, and TKE distributions. While appealing, Ayik's model is
nevertheless phenomenological in nature (as is random matrix theory).
It has been applied recently to the case of the spontaneous fission of
$^{258}$Fm~\cite{Tanimura} in an approach characterized by the authors somewhat
optimistically as ``the first time a fully microscopic description of the fragment TKE
distribution after fission and gives unique microscopic information on the fission process.''
The main difference of the approach of Ref.~\cite{Tanimura} and TDHF calculations
performed in the past is in choosing an {\it ad hoc} ensemble of stochastic initial conditions,
instead of a single initial nuclear shape past the outer fission barrier. In the case of
spontaneous fission  the nucleus emerges from under the barrier likely in its
ground state corresponding to that instantaneous shape with $Q_{20}\approx 100$ barn for
$^{258}$Fm~\cite{Tanimura}. These authors~\cite{Tanimura}  however
chose to start their fission dynamics much further, when $Q_{20}$ is either 125 or 160 barn,
at which point they excite their nucleus to an energy
corresponding to the ground state. This stochastic ensemble has
fluctuations of all possible observables $(N, Z, E)$, center-of-mass velocity,
intrinsic currents, total angular momentum, parity, etc. of uncharacterized distributions.
\begin{wrapfigure}{r}{0.57\textwidth}
\includegraphics[clip,width=0.57\textwidth]{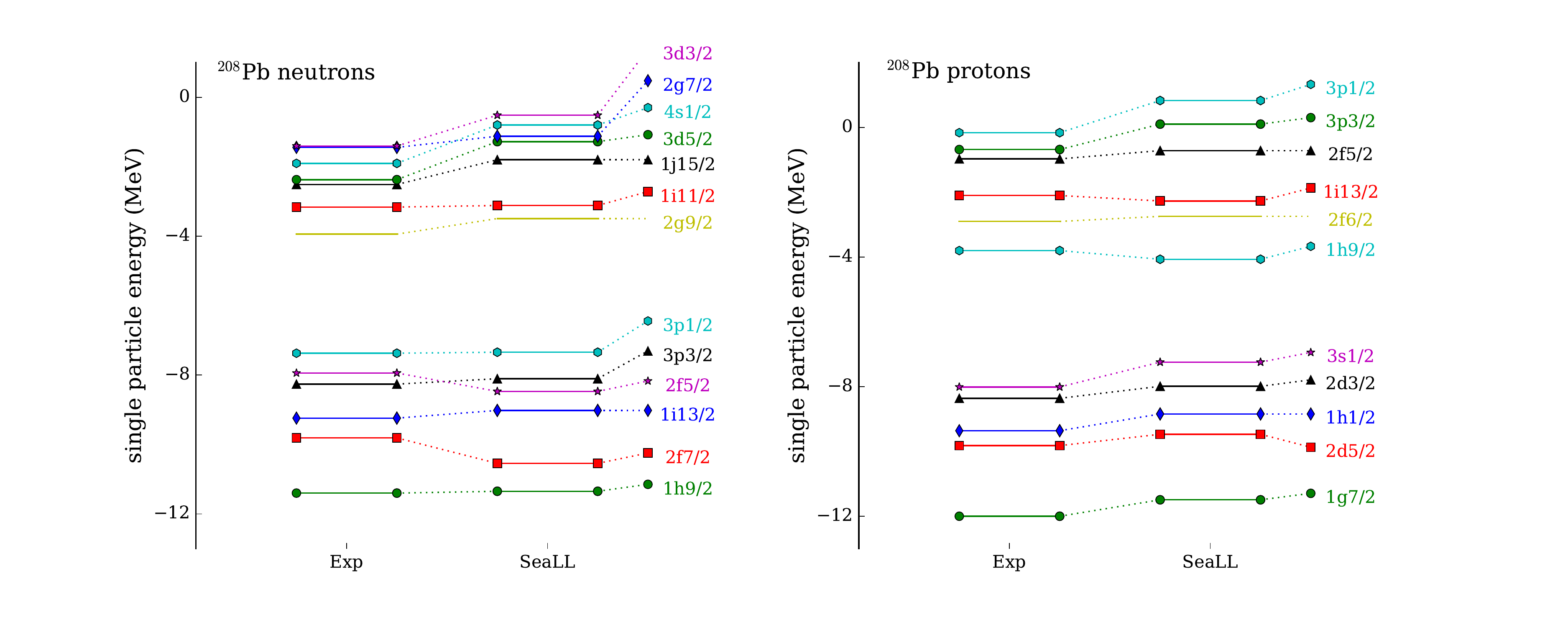}
\caption{ \label{fig:ab6}   The single-particle energy levels  $^{208}$Pb evaluated with
the new energy density functional SeaLL1~\cite{NEDF} versus experiement.}
\end{wrapfigure}
When the nucleus exits from under the barrier
it is expect to start rolling down towards scission configurations rather slowly and if
intrinsic degrees of freedom get excited, one naturally would expect a thermal
ensemble for the nuclear shapes with $Q_{20}$ = 125 or 160 barn,
rather that the stochastic ensemble suggested by Ayik~\cite{Ayik}.
There is no theoretical argument presented in Ref.~\cite{Tanimura} to choose as a starting point
of the dynamical simulations the arbitrary deformations  $Q_{20}$ = 125 or
160 barn, by which time the nucleus acquired a non-negligible collective
kinetic energy not explicitly accounted for in this approach. It would
be more natural to start the simulation at the exact
configuration where the nucleus emerged from under the barrier
at $Q_{20}\approx$ 100 barn. In that case the size of the ``quantum
fluctuations'' would be zero and the nucleus would have no intrinsic
excitation energy and zero collective momentum.
But that would deprive the authors from the ability to generate
a FFs distribution. The authors even establish that if they were to start their simulations closer to
the scission configurations their results would be quite different, thus precluding this approach
of its predictive power.  These authors conclusion that simulations with and without ``dynamical pairing"
lead to similar results also appear quite hasty, as in both cases one starts with
an ensemble of single Slater determinants, which probably share some
similar fluctuations. One can safely conclude that the FFs
distribution definitely depend strongly  on the initial conditions within such a stochastic
approach. A more natural approach would be a TDGCM approach advocated by
Regnier {\it et al.}~\cite{Regnier} to specify the distribution of the
initial conditions in a TDDFT simulation of fission dynamics.

The predictive power of a TDDFT approach depends critically on the
quality of the NEDFs, which are phenomenological so far and depend on a
rather large number of fitting parameters (typically 14 and sometimes even more),
some of them rather strongly correlated with each other. This points to the fact
that there are elements of NEDF that are still poorly understood
and there is a need for better constrained NEDFs, either phenomenologically or microscopically.
We have developed a qualitatively
new NEDF~\cite{NEDF}, which contains terms previously not considered in literature,
which depends on only 7 uncorrelated parameters, with a rms for masses of even-even nuclei
from the AME2012~\cite{ame2012} of 1.74~MeV and for charge radii of 0.034~fm.
The potential energy surface for $^{240}$Pu predicted by SeaLL1, see Figure~\ref{fig:ab5},
has potential energy barriers comparable to those obtained in FRLDM~\cite{LDM},
though a relatively low energy of the fission isomer, and a nice single-particle spectrum of
$^{208}$Pb, see Figure~\ref{fig:ab6}.

This work was supported in part by U.S. Department of Energy (DOE)
Grant No. DE-FG02-97ER41014 and by the U.S. Department of Energy under Contract Nos.\
DE-AC52-07NA27344 (Lawrence Livermore National Laboratory).
This research used resources of the Oak Ridge Leadership Computing Facility, which is a DOE Office of Science User Facility supported under Contract DE-AC05-00OR22725.
It also used resources of the National Energy Research Scientific Computing Center, a DOE Office of Science User Facility supported by the Office of Science of the U.S. Department of Energy under Contract No. DE-AC02-05CH11231.
Some of the calculations were performed on Moonlight at LANL.
Computing support for this work also came from the Lawrence Livermore National
Laboratory (LLNL) Institutional Computing Grand Challenge program.

\end{document}